\documentclass[aps,pra,twocolumn,amsmath,showpacs,letterpaper,floatfix]{revtex4}

\newcommand{\bra}[1]{\left\langle #1\right|}
\newcommand{\ket}[1]{\left| #1\right\rangle}

\newcommand{\eeqref}[1]{Eq.~(\ref{#1})}

\usepackage{amssymb,amsmath,amsthm,amsfonts}
\usepackage{graphicx}
\usepackage{epsfig}
\usepackage{subfigure}
\usepackage{amsmath}			
\usepackage{color}

\def\4F3{\mbox{$_4${F}$_3$}}
\def\t4F3{\mbox{$_4{\tilde {\rm F}}_3$}}

\begin{document}

\title{Moments of nonclassicality quasiprobabilities}

\author{Saleh Rahimi-Keshari$^{1,2}$, Thomas Kiesel$^{1}$, and Werner Vogel$^{1}$}
\affiliation{$^{1}$Arbeitsgruppe Quantenoptik, Institut f\"ur Physik, Universit\"at Rostock, D-18051 Rostock, Germany \\
$^{2}$Centre for Quantum Computation and Communication Technology,
School of Mathematics and Physics,
University of Queensland, St Lucia, Queensland 4072, Australia}
%\email{}

%opening
\begin{abstract}

A method is introduced for the verification of nonclassicality in terms of moments of nonclassicality quasiprobability distributions. The latter are
easily obtained from experimental data and will be denoted as nonclassicality moments. Their relation to  normally-ordered moments is derived, which enables us to verify nonclassicality by using well established criteria. Alternatively, nonclassicality criteria are directly formulated in terms of nonclassicality moments. The latter converge in proper limits to the usually used criteria, as is illustrated for squeezing and sub-Poissonian photon statistics.
Our theory  also yields expectation values of any observable in terms of nonclassicality moments.
\end{abstract}

% TK: Candidates for PACS
% 02.50.Cw 	Probability theory 
% 02.50.Fz 	Stochastic analysis
% 03.65.Ta 	Foundations of quantum mechanics; measurement theory (for optical tests of quantum theory, see 42.50.Xa)
% 03.65.Wj 	State reconstruction, quantum tomography 
% 42.50.Ar 	Photon statistics and coherence theory 
% 42.50.Xa 	Optical tests of quantum theory 
\maketitle

\section{Introduction}\label{intro}

The rapidly developing experimental techniques opened new fields of research
which make use of the basic principles of quantum physics. Beyond the experimental demonstration of quantum phenomena, nowadays new types of quantum technologies are 
aimed at being developed. In this context there arises a renewed interest in the characterization of the quantum properties of light and matter. A clear characterization and interpretation of quantum phenomena, including the quantum interference effects, play a central role for beating the technical limitations known in classical physics.

In the field of quantum optics the characterization of quantum effects of light was based on the Glauber-Sudarshan $P$~representation of the density operator~\cite{Sudarshan,Glauber},
\begin{equation}
\rho=\int \text{d}^2\alpha P(\alpha,\alpha^*) \ket{\alpha}\bra{\alpha}.
\label{Glauber-Sudarshan}
\end{equation}
In this form, the density operator $\rho$ of a single mode radiation field is expressed as a formal pseudo-mixture of the coherent states $\ket{\alpha}$.
The latter are those quantum states of the harmonic oscillator which are closest to its classical behavior. Whenever $P(\alpha,\alpha^*)$ has the properties of a classical probability density, the corresponding quantum state is a classical
mixture of the (in our terms classical) coherent states. Such quantum states have been addressed as those having a classical analog~\cite{Titulaer}.
Whenever $P$ fails to be a probability density, in particular if it has negativities,
the quantum state is said to be a nonclassical one~\cite{Mandel}. Nonclassicality of this type is  indispensable for the occurrence of quantum interferences, which play the key role for most of the presently considered applications of quantum physics.

It is important to note that this traditional characterization of nonclassical states is not limited to the consideration of single-mode systems. The extension to multi-mode scenarios is straightforward. It only requires to replace the coherent amplitude $\alpha$ by a vector, whose components $\alpha_i$ ($i=1,\dots, N$) describe  an $N$-mode system. In such a description the nonclassicality also includes -- as special cases -- such important phenomena like entanglement, for recent reviews we refer to~\cite{Horodecki,Guehne,Ferry}. More generally, by introducing a
$P$~functional, one may even characterize space-time dependent radiation field correlation properties~\cite{Vogel}, which even include the dynamics of the light-emitting radiation sources.

Despite the usefulness of the $P$~function for describing quantum phenomena,
it has some severe deficiencies for practical applications. Most importantly,
it not only may attain negative values, but in general it is a strongly singular distribution which is not accessible by the methods of quantum-state reconstruction, for reviews see~\cite{Leonhard,Welsch,Raymer}. To overcome such problems, the nonclassicality criteria have been reformulated in terms of characteristic functions~\cite{Vogel-cf,Ri-Vo}, for applications in experiments see~\cite{Lvovsky,Zava,Kie-Schn}.
In general, however, such methods require to analyze hierarchies of nonclassicality conditions.
 
An alternative approach is the possibility to regularize the $P$~function in some form~\cite{Klauder,Cahill}.
On this basis, however, the characterization of nonclassical effects becomes a complicated issue in general. For this purpose, recently a regularization of the $P$~function has been proposed, which directly displays the nonclassicality as negativities of a so-called nonclassicality quasiprobability (NQP)~\cite{Kiesel}. This distribution can be experimentally determined in a simple manner, from the data recorded by balanced homodyning~\cite{Kiesel-POmega-Spats}. Assuming the NQP has been determined by direct sampling, for details see~\cite{Kiesel-POmega-Squeeze},
the general information about the quantum state and its nonclassical effects is available in principle. Nevertheless, the question arises how one can use the NQP to calculate physical expectation values. This requires knowledge of the relation between the moments of the NQP -- in the following denoted as nonclassicality moments -- to other moments, such as the normally-ordered ones obtained from the $P$~function.

The aim of the present paper is to provide methods for expressing general expectation values via NQPs and to characterize quantum effects by nonclassicality moments. In Sec.~II we deal with the NQPs and their moments. The relations between nonclassicality moments and normally-ordered moments is considered in Sec.~III. 
Examples of nonclassical effects and particular expectation values are studied in Sec.~IV. A summary and some conclusions are given in Sec.~V.

\section{Nonclassicality quasiprobabilities and moments}\label{NQP}

% A universal method for verification of nonclassicality of unknown quantum states has been of great practical interest, specially due to advent of quantum information science. A single-mode quantum optical state is referred to as nonclassical if and only if the $P$ function in its Glauber-Sudarshan representation \cite{Glauber,Sudarshan}
% \begin{equation}
% \rho=\int \text{d}^2\alpha P(\alpha,\alpha^*) \ket{\alpha}\bra{\alpha},
% \label{Glauber-Sudarshan}
% \end{equation}
% violates the properties of a probability density. However, as for many quantum states the $P$ function is a highly singular function, nonclassicality of a quantum state, in general, cannot be verified by measuring the $P$ function.  
% Hence, a practical method must be invoked.   

%Recently, a new class of quasiprobability distributions was introduced so that their negativity is an indicator of nonclassicality of any %quantum state.  Hence it is referred to as the nonclassicality quasiprobability (NQP) distribution~\cite{Kiesel},
%which we describe by the symbol $P_{\Omega}(\alpha,\alpha^*)$. 
The characteristic function of the NQP is obtained by multiplying the characteristic function of the $P$ function by a filter function $\Omega_{w}(\xi)$, 
\begin{equation}
\Phi_{\Omega}(\xi,\xi^*)= \Phi(\xi,\xi^*) \Omega_{w}(\xi,\xi^*)
\label{charac-filter}
\end{equation}
where $w$ is a real parameter controlling the width of the filter and $\Omega_{w}(\xi,\xi^*)$ must satisfy certain conditions: $\Omega_w(\xi,\xi^*)e^{|\xi|^2/2}$ is square-integrable; its Fourier transform is a positive-semidefinite function~\cite{Kiesel}. Also, in order that the relation \eqref{charac-filter} be invertible, it is required that the filter has no zeros \cite{Ag-Wo}.  
It is assumed that $\Omega_{w}(0)=1$, and the dependence on $w$ of the filter is introduced as a scaling factor $\Omega_{w}(\xi,\xi^*)=\Omega\left(\frac{\xi}{w},\frac{\xi^*}{w}\right)$ so that $\lim_{w\rightarrow\infty} \Omega_{w}(\xi,\xi^*)= 1 $. The advantage of using NQP, which denoted by the symbol $P_{\Omega}(\alpha,\alpha^*)$, to test nonclassicality is that it is a regular function that can be directly obtained by balanced homodyne detection, even via direct sampling~\cite{Kiesel-POmega-Squeeze}.
Furthermore, it has been proved that for all nonclassical states one may find a width $w$ and a point $\alpha_{0}$ such that $P_\Omega(\alpha_{0},\alpha_{0}^*)$ is negative. However, for some nonclassical states a large value of the width parameter $w$ is required to observe the negativity of NQP such that the inherent statistical uncertainties due to experimental measurement may hide all nonclassical effects. 

% \SRK{In the next section, we derive a relation between the the experimentally accessible moments of NQP
% \begin{equation}
% \label{nonclass-mom}
%  M_{\Omega, nm}=\int \text{d}^2\alpha P_{\Omega}(\alpha,\alpha^*) {\alpha^*}^n {\alpha}^m \ , 
% \end{equation} 
% which are referred to as {\it nonclassicality moments} of a quantum state, and the normally-ordered moments of a quantum state. \TK{%This is equivalent to calculate the moments for infinite $w$, but does not require the reconstruction of the corresponding quasiprobability. 
% This enables us 
% to verify nonclassicality based on normally-ordered moments obtained from the nonclassicality moments for a given value of the width $w$, and calculate the expectation value of any observable in terms of the nonclassicality moments.}
% In order that the nonclassicality moments being well-defined, in addition to the above-mentioned conditions, we must require that the filter $\Omega_{w}(\xi,\zeta)$ be an entire function of two complex variables $\xi$ and $\zeta$.}

{The aim of this paper is to provide a method for verifying nonclassicality properties of quantum states based on using the moments of the experimentally accessible NQP
\begin{equation}
\label{nonclass-mom}
 M_{\Omega, nm}=\int \text{d}^2\alpha P_{\Omega}(\alpha,\alpha^*) {\alpha^*}^n {\alpha}^m \ , 
\end{equation}  
which are referred to as {\it nonclassicality moments} of a quantum state. In order that the nonclassicality moments being well-defined, in addition to the above-mentioned conditions, we must require that the filter $\Omega_{w}(\xi,\zeta)$ be an entire function of two complex variables $\xi$ and $\zeta$.} % \cite{new-pap}. }
{We also may consider the nonclassicality moments as the normally-ordered moments of an operator,}
\begin{equation}
\tilde{\rho} = \int \text{d}^2\alpha P_{\Omega}(\alpha,\alpha^*) \ket{\alpha}\bra{\alpha} \ ,
\label{filtered-state}
\end{equation}
so that
\begin{equation}
\label{defi}
 M_{\Omega, nm}=\langle{a^{\dagger}}^n a^m\rangle_{\tilde{\rho}}.
\end{equation}
If one observes nonclassical effects, such as photon-antibunching~\cite{Kimble},
sub-Poissonian statistics~\cite{Short}, and squeezing~\cite{Slusher}, in terms of
nonclassicality moments, the NQP $P_{\Omega}(\alpha,\alpha^*)$ in \eeqref{filtered-state} must attain negative values, and hence the quantum state $\tilde \rho$ must be nonclassical. Therefore, all moment criteria, such as the nonnegativity of different matrices of moments~\cite{AgarwalTara,Agarwal-xmoment,Shchukin-Vogel,Sh-Vo,Miranowicz}, can be applied to verify nonclassicality, instead of seeking negativity in the NQP.

%%%%%%%%%%%%%%%%%%%%%%%%%%%%%%%%%%%%%%%%%%%%%%%%%%%%%%%%%%%%%%%%%%%%%%%%%%%%%%%%%%%%%%%%%%%%%%%%%%%%%%%%%%%%%%%%%%%%%%%%%%%%%%%%%%%%%%%%%%%%%%%
\section{Nonclassicality moments and normally-ordered moments}\label{filter-map}

{In the following, we derive a relation between the nonclassicality moments and the normally-ordered moments of a quantum state. This relation enables us 
to verify nonclassicality based on normally-ordered moments obtained from the nonclassicality moments for a given value of the width $w$.
This is equivalent to calculate the moments for infinite $w$, but does not require the reconstruction of the corresponding quasiprobability.
Also, by using this relation one can calculate the expectation value of any observable in terms of the nonclassicality moments. 
We show that in the limiting case of large values of $w$ the nonclassicality moments converge to normally-ordered ones.}

\subsection{Relation between nonclassicality and normally-ordered moments}

The NQP is a representation of a quantum state, and, in principle, has all the information about the state; hence, if all the nonclassicality moments exist, they uniquely determine the quantum state. In this section, we derive a relation between nonclassicality and normally-ordered moments by which all known criteria for nonclassicality that are based on normally-ordered moments can be related to the nonclassicality moments.

By using~\eeqref{nonclass-mom} and expressing $P_{\Omega}(\alpha,\alpha^*)$ in terms of its Fourier transform, the nonclassicality moments are given by
\begin{align}
 M_{\Omega,nm}&= \int \text{d}^2\xi \Phi(\xi,\xi^*) \Omega(\xi,\xi^*) \int \frac{\text{d}^2\alpha}{\pi^2}  e^{\alpha\xi^*-\xi\alpha^*} {\alpha^*}^n {\alpha}^m \nonumber \\
&= \int \text{d}^2\xi \Phi(\xi,\xi^*) \Omega(\xi,\xi^*) (-1)^n \frac{\partial^n}{\partial\xi^n} \frac{\partial^m}{\partial{\xi^*}^m} \delta^{2}(\xi)  \nonumber \\
&=(-1)^m \frac{\partial^n}{\partial\xi^n} \frac{\partial^m}{\partial{\xi^*}^m}\Phi(\xi,\xi^*) \Omega(\xi,\xi^*)\Big|_{\xi=0} \ .
\label{map-mom}
\end{align}
By applying the relations
\begin{equation}
 \frac{\text{d}^n}{\text{d}x^n} f(x) g(x) = \sum_{i=0}^{n} \binom{n}{i} \frac{\text{d}^{n-i}}{\text{d}x^{n-i}}f(x) \frac{\text{d}^i}{\text{d}x^i} g(x) \ ,
\end{equation}
and 
\begin{align}
\langle{a^{\dagger}}^n a^m\rangle &= \int \text{d}^2\alpha P(\alpha,\alpha^*)  {\alpha^*}^n {\alpha}^m \nonumber \\
&= (-1)^m \frac{\partial^n}{\partial\xi^n} \frac{\partial^m}{\partial{\xi^*}^m}\Phi(\xi,\xi^*) \Big|_{\xi=0} \ ,
\end{align}
we express the nonclassicality moments in terms of the normally-ordered ones,
\begin{equation}
 M_{\Omega,nm} = \sum_{i=0}^{n} \sum_{j=0}^{m} (-1)^j C_{i,j} \binom{n}{i} \binom{m}{j} \langle{a^{\dagger}}^{n-i} a^{m-j}\rangle \ ,
\label{mom-rel-nonclas-normal}
\end{equation}
with
\begin{equation}
 C_{i,j} = \frac{\partial^i}{\partial\xi^i} \frac{\partial^j}{\partial{\xi^*}^j} \Omega(\xi,\xi^*)\Big|_{\xi=0} \ .
 \label{coef}
\end{equation}
 
Note that, similarly, one can express the normally-ordered moments in terms of the nonclassicality moments
\begin{equation}
\label{mom-rel-normal-nonclas}
\langle{a^{\dagger}}^{n} a^{m}\rangle = \sum_{i=0}^{n} \sum_{j=0}^{m} (-1)^j \bar{C}_{i,j} \binom{n}{i} \binom{m}{j} M_{\Omega,n-i,m-j}
\end{equation}
with
\begin{equation}
 \bar{C}_{i,j} = \frac{\partial^i}{\partial\xi^i} \frac{\partial^j}{\partial{\xi^*}^j} \Omega(\xi,\xi^*)^{-1}\Big|_{\xi=0} \ .
 \label{coef2}
\end{equation}
Therefore, according to equations \eqref{mom-rel-nonclas-normal} and \eqref{mom-rel-normal-nonclas}, by calculating the coefficients $C_{i,j}$ and $\bar{C}_{i,j}$, nonclassicality moments and normally-ordered moments can be expressed in terms of each other.

The filter function, which was used in Ref. \cite{Kiesel}, is the autocorrelation of $\exp(-|\xi|^4)$, 
\begin{equation}
\Omega\left(\frac{\xi}{w},\frac{\xi^*}{w}\right)=\frac{1}{N} \int \text{d}^2 \beta e^{-|\beta|^4} e^{-|\frac{\xi}{w} + \beta|^4} \ ,
\label{nonclas-filter}
\end{equation}
where $N=\int \text{d}^2 \beta e^{-2|\beta|^4} $. We use this filter function to obtain the coefficients in the moments relations \eqref{mom-rel-nonclas-normal} and \eqref{mom-rel-normal-nonclas}. For this purpose, we apply Eq.~\eqref{coef} for the nonclassicality filter. 
By changing the variable $u=\xi /w$, and using
\begin{equation}
\frac{\partial^n}{\partial\xi^n}=\frac{1}{w^n} \frac{\partial^n}{\partial u^n} \ ,
\end{equation} 
\eeqref{coef} becomes
\begin{equation}
C_{i,j} =\frac{C'_{i,j}}{w^{i+j}} 
\end{equation}
with
\begin{equation}
C'_{i,j}=\frac{\partial^i}{\partial u^i} \frac{\partial^j}{\partial{u^*}^j} \Omega(u,u^*)\Big|_{u=0} \ .
\label{coef-nonclas}
\end{equation}
For the filter function \eqref{nonclas-filter} we can calculate the coefficients \eqref{coef-nonclas}, which yields
\begin{equation}
C'_{i,j}=\delta_{i,j}\ \sqrt{2\pi} 2^{2i}   \t4F3(\frac{1}{2},\frac{1}{2},1,1;\,1-\frac{i}{2},1-\frac{i}{2},\frac{1-i}{2};\, -1),
\label{coef-non}
\end{equation}
with
\begin{align}
\t4F3(a_1,a_2,a_3,a_4;\,b_1,&b_2,b_3;\, x)= \nonumber \\
& \frac{\4F3(a_1,a_2,a_3,a_4;\,b_1,b_2,b_3;\, x)}{\Gamma(b_1)\Gamma(b_2)\Gamma(b_3)}
\end{align}
%$\t4F3(a_1,a_2,a_3,a_4;\,b_1,b_2,b_3;\, x) = \4F3(a_1,a_2,a_3,a_4;\,b_1,b_2,b_3;\, x)/(\Gamma(b_1)\Gamma(b_2)\Gamma(b_3))$ }
being the regularized generalized hypergeometric function. 

Using Eq.~\eqref{coef-non}, the nonclassicality moments can be expressed in terms of the normally-ordered moments as
\begin{align}
M_{\Omega,nm} &=  \sum_{i=0}^{\min(m,n)} (-1)^i \langle{a^{\dagger}}^{n-i}a^{m-i}\rangle \binom{n}{i} \binom{m}{i} \frac{\sqrt{2\pi} 2^{2i}}{w^{2i}} \nonumber \\ 
&\times \t4F3(\frac{1}{2},\frac{1}{2},1,1;\,1-\frac{i}{2},1-\frac{i}{2},\frac{1-i}{2};\, -1) \ .  
\label{nonclas-normal-rel}
\end{align}
Conversely, the normally-ordered moments can be found in terms of the nonclassicality moments by using Eq.~\eqref{mom-rel-normal-nonclas}. Alternatively, based on \eeqref{nonclas-normal-rel} we obtain a set of linear equations which relate the two types of moments to each other. The solution of these equations also yields the sought inverse relation for the normally-ordered moments.
For example, from \eeqref{nonclas-normal-rel} we readily obtain
\begin{equation}
\langle{a^{\dagger}}^n\rangle=M_{\Omega,n0}\ , \ \ \ \ \ 
\langle a^{m}\rangle=M_{\Omega,0m}\ , 
\end{equation}
\begin{equation}
\label{mom11}
\langle a^\dagger a\rangle = M_{\Omega,11} - \sqrt{\frac{2}{\pi}} \frac{1}{w^2},\end{equation}
and
\begin{equation}
\label{mom22}
\langle{a^{\dagger}}^{2} a^2\rangle = M_{\Omega,22} - \sqrt{\frac{2}{\pi}}\frac{4}{w^2} M_{\Omega,11} + \left(\frac{8}{\pi}-\frac{7}{4}\right)\frac{1}{w^4}.
\end{equation}
These relations will be used in the following for expressing nonclassicality criteria in terms of nonclassicality moments.

By expressing the normally-ordered moments of a quantum state in terms of the nonclassicality moments, one may calculate the expectation value of any observable $B$, given in the form
\begin{equation}\label{observ}
B=\sum_{n,m=0}^{\infty} b_{nm} {a^{\dagger}}^{n} a^{m} \ .
\end{equation}
Applying \eeqref{mom-rel-normal-nonclas}, the expectation value of the observable
reads as
\begin{align}\label{observable}
\text{Tr}[\rho B] &= \sum_{n,m=0}^{\infty} b_{nm} \langle{a^{\dagger}}^{n}a^{m}\rangle \nonumber \\
&= \sum_{n,m=0}^{\infty} b_{nm} \sum_{i=0}^{n} \sum_{j=0}^{m} (-1)^j \bar{C}_{i,j} \binom{n}{i} \nonumber \\
&\times\binom{m}{j} M_{\Omega,n-i,m-j} \ .
\end{align}
The nonclassicality moments occurring in this expression are easily obtained, provided that the NQP has been experimentally determined through direct sampling. If one would express the observables in terms of normally-ordered moments, as in the first line of \eeqref{observable}, the latter must be obtained by methods of state reconstruction~\cite{Richter-mom,Manko} or by homodyne correlation measurements~\cite{Sh-Vo-corr}.

%%%%%%%%%%%%%%%%%%%%%%%%%%%%%%%%%%%%%%%%%%%%%%%%%%%%%%%%%%%%%%%%%%%%%
\subsection{Convergence of nonclassicality moments}

An interesting feature of \eeqref{nonclas-normal-rel} is that for large values of the width parameter $w$ the moments of the NQP converge to the normally-ordered moments. For $w \gg 1$ and $n,m \neq 0$, and using \eqref{defi}, we have 
\begin{equation}
\left\lvert \langle{a^{\dagger}}^{n} a^{m}\rangle_{\tilde{\rho}} - \langle{a^{\dagger}}^{n} a^{m}\rangle_{\rho} \right\rvert = O(\frac{1}{w^2}) \ .
\label{con-mom}
\end{equation}
Therefore, for sufficiently large values of $w$ the normally-ordered moments of $\tilde{\rho}$ can be considered as a good approximation of those of the original quantum state $\rho$.

Moreover, \eeqref{con-mom} implies that for any observable operator $B$, 
we have
\begin{align}
\left\lvert \text{Tr}[\rho B] -\text{Tr}[\tilde{\rho} B] \right\rvert &= \left\lvert \sum_{n,m=1}^{\infty} b_{nm} \left( \langle {a^{\dagger}}^{n} a^{m}\rangle_{\tilde{\rho}} - \langle{a^{\dagger}}^{n} a^{m}\rangle_{\rho} \right) \right\rvert \nonumber \\
&\leqslant \sum_{n,m=1}^{\infty} \left\lvert b_{nm} \left( \langle{a^{\dagger}}^{n} a^{m}\rangle_{\tilde{\rho}} - \langle{a^{\dagger}}^{n} a^{m}\rangle_{\rho} \right) \right\rvert \nonumber \\
&=O(\frac{1}{w^2}) .
\end{align}
Therefore, the expectation value of the observable $B$ can be approximated with arbitrary degree of accuracy by $\text{Tr}[\tilde{\rho} B]$, provided that one may choose a sufficiently large value of $w$. In practise, however, this possibility is limited by the statistical noise in the available set of experimental data.

%%%%%%%%%%%%%%%%%%%%%%%%%%%%%%%%%%%%%%%%%%%%%%%%%%%%%%%%%%%%%%%%%%%%%%%%%%%%%%%%%%%%%%%%%%%%%%%%%%%%%%%%%%%%%%%%%%%%%%%%%%%%%%%%%%%%%%%%%%%%%%%
\section{Nonclassical effects in terms of nonclassicality moments}\label{examples}

It is known that the function $P_\Omega(\alpha,\alpha^*)$ can be obtained from balanced homodyne detection and quantum state reconstruction~\cite{Kiesel-POmega-Spats,Kiesel-POmega-Squeeze}. 
Therefore, one can obtain the moments of $P_\Omega(\alpha,\alpha^*)$ from experimental data. Since $P_\Omega(\alpha,\alpha^*)$ has been designed for the verification of nonclassicality, we may look for nonclassical effects in the corresponding nonclassicality moments.

As examples of nonclassical effects, in the following we consider sub-Poissonian photon statistics and squeezing. We show that for sufficiently large values of $w$ the Mandel-Q-parameter and the quadrature variance in terms of the nonclassicality moments exhibit the corresponding nonclassical effects. Alternatively, by using the derived relation between the nonclassicality moments and the normally-ordered moments, one can verify the nonclassical effects in terms of the normally-ordered moments obtained from the nonclassicality moments for a given value of $w$.

%%%%%%%%%%%%%%%%%%%%%%%%%%%%%%%%%%%%%%%%%%%%%%%%%%%%%%%%%%%%%%%%%%%%
\subsection{Sub-Poissonian statistics}

Let us start to examine nonclassical effects of a single-photon added thermal state (SPATS)~\cite{AgarwalTara}. The NQP of a SPATS has been reconstructed to verify its nonclassical behavior~\cite{Kiesel-POmega-Spats}. Since these states are diagonal in Fock basis, we look at moments of the photon number. Taking Eq.~(\ref{nonclas-normal-rel}) and inserting the moments for SPATS explicitly, we find
\begin{align}
	M_{\Omega,11} & =  2\bar n + 1+ \sqrt{\frac{2}{\pi}}\frac{1}{w^2},\\
	M_{\Omega,22} & =  6\bar{n}^2 + 4 \bar n
 + \sqrt{\frac{2}{\pi}}\frac{4}{w^2}\left(2\bar n+1\right) + \frac{7}{4 w^4},
\end{align}
where $\bar{n}$ denotes the mean thermal photon number of the SPATS under study.
With these moments, we can introduce the Mandel-Q-parameter~\cite{MandelQ} of the operator $\tilde\rho$ defined in \eeqref{filtered-state},
\begin{equation}
	Q_\Omega = \frac{M_{\Omega,22} - M_{\Omega,11}^2}{M_{\Omega,11}}.
\end{equation}
If this quantity is negative, then we can conclude negativities in the nonclassicality quasiprobability $P_\Omega(\alpha,\alpha^*)$. Due to the properties of the nonclassicality filters, the original state $\rho$ must be nonclassical as well. For $\tilde\rho$, we find
\begin{equation}
	Q_{\Omega} = \frac{7\pi - 8 + 8(2 \bar n+1) \sqrt{2 \pi } w^2+\pi\left(8 \bar n^2-4\right) w^4}{4 w^2 \left(\sqrt{2 \pi }+(2 \bar n+1) \pi  w^2\right)}
\end{equation}
In the limit $w\to\infty$, we obtain  the $Q$ parameter of the original (unfiltered) SPATS,
\begin{equation}
	Q\equiv\lim_{w\to\infty} Q_{\Omega} = \frac{2 \bar n^2-1}{2 \bar n+1},\label{eq:Mandel:Q}
\end{equation}
which has also been considered in~\cite{AgarwalTara}. Therefore, we may expect negative values of the original $Q$~parameter if the mean thermal photon number satisfies the condition $\bar n < \sqrt{2}/2$. 

\begin{figure}[tpb]
\centering
	\includegraphics[width=0.9\columnwidth]{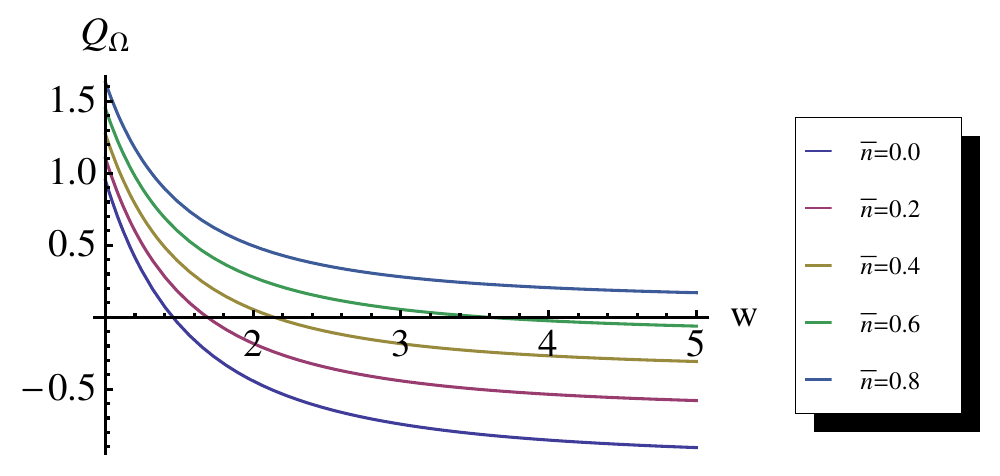}
	\caption{Dependence of the $Q_\Omega$ parameter on the filter width.}
	\label{fig:Q}
\end{figure}

Figure~\ref{fig:Q} shows the dependence of $Q_\Omega$ on the filter width $w$, for different mean thermal photon numbers $\bar n$. The larger the filter width, the larger the negativity of $Q_\Omega$ becomes.  Moreover, increasing the mean thermal photon number leads to decreasing negativities in $Q_\Omega$. For $\bar n = 0.8$, nonclassical effects cannot be seen by means of $Q_\Omega$, as it is expected from Eq.~(\ref{eq:Mandel:Q}). 
	
\begin{figure}[tpb]
\centering
	\includegraphics[width=0.8\columnwidth]{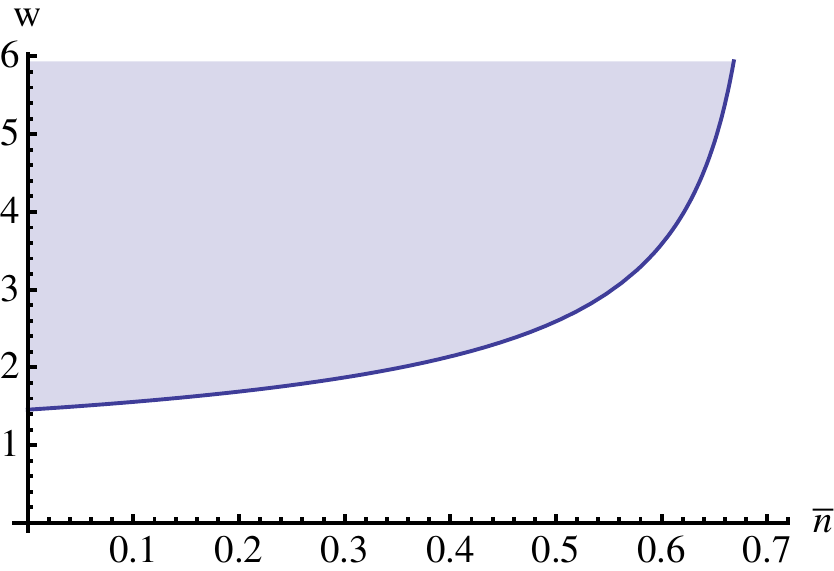}
	\caption{Filter widths $w$ which lead to a negative $Q$ parameter for different mean thermal photon numbers.}
	\label{fig:w0}
\end{figure}

Furthermore, we can look for the minimum filter width $w_0$, for which negativities of $Q_\Omega$ appear. 
 The blue shaded area in Fig.~\ref{fig:w0} indicates the possible filter widths $w$ for which $Q_\Omega$ is negative. The blue boundary is the set of $w$ for which the negativities of the Mandel-Q parameter vanish. We observe that a larger filter width is required in order to detect nonclassicality when the
mean thermal photon number is increasing.

% \begin{align}
% 	w_0^2 = \frac{-2 \sqrt{2} (1+2 \bar n)+\sqrt{2 \bar n (16+\bar n (24-7 \pi ))+7 \pi }}{2 \left(2 \bar n^2-1\right) \sqrt{\pi }}
% \end{align}

As discussed above, it may happen that it is not possible to detect negativities in $P_\Omega(\alpha,\alpha^*)$ for an accessible range of the width $w$. However, one could reconstruct the moments for a lower filter width $w$ and invert Eq.~\eqref{nonclas-normal-rel} in order to estimate the normally-ordered moments. 
For instance, one may use the first and second normally-ordered moment of the photon number, Eqs. \eqref{mom11} and \eqref{mom22}, which are required for the Mandel-Q parameter. Thus one can estimate the $Q$ parameter from measured moments of $\tilde\rho$. In this manner one may detect nonclassicality of states, for which a large filter width of the nonclassicality filter would be required.
For example, from Fig.~\ref{fig:Q} it is seen that the $Q_\Omega$ parameter is positive for $w<3.5$ and $\bar{n}=0.6$ , while the standard $Q$ parameter turns out to be negative.

%neu%%%%%%%%%%%%%%%%%%%%%%%%%%%%%%%%%%%%%

\subsection{Squeezing}

\begin{figure}
     \includegraphics[width=0.8\columnwidth]{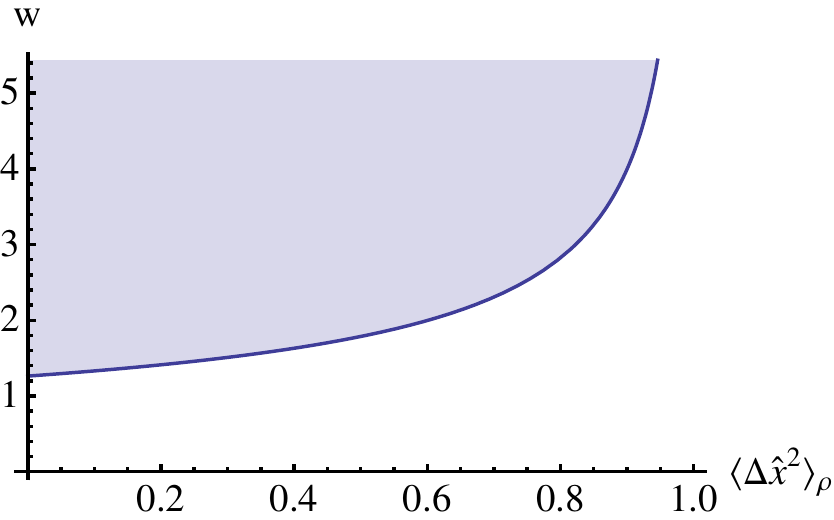}
     \caption{Filter widths $w$ which lead to squeezing of the filtered 
quadrature variance $\langle \Delta x^2 \rangle_{\tilde \rho}$, in 
dependence of the standard quadrature variance $\langle \Delta x^2 
\rangle_{\rho}$.}
     \label{fig:w0:squeeze}
\end{figure}

In order to detect the squeezing effect based on the nonclassicality 
moments, using \eeqref{mom11}, we obtain
\begin{equation}
\label{sq-var}
\langle \Delta x^2 \rangle_{\tilde \rho} = \langle \Delta x^2 
\rangle_{\rho} + 2\sqrt{\frac{2}{\pi}}\frac{1}{w^2}.
\end{equation}
Thus, for any amount of squeezing, when
\begin{equation}
     \langle \Delta x^2 \rangle_{\rho} < 1,
\end{equation}
one can choose sufficiently large values of the width $w$ such that the 
variance of the quadrature operators of $\tilde\rho$ exhibit squeezing. 
In Fig.~\ref{fig:w0:squeeze}, we show the minimum required filter width 
$w$ for observing squeezing for the filtered quasiprobability. The 
larger the squeezing effect, the less is the minimum width. We note that 
even for infinite squeezing, $\langle \Delta x^2 \rangle_{\rho} = 0$, a 
finite width $w$ is needed to observe the desired effect.

Alternatively, the squeezing effect can be verified by calculating the 
quadrature variance from \eqref{sq-var}. One only has to 
invert the equation, which can be trivially done.
This immediately yields the squeezing condition in terms of nonclassicality moments as
\begin{equation}
\label{sq-noncl-mom}
\langle \Delta x^2 \rangle_{\tilde \rho} < 1 + 
2\sqrt{\frac{2}{\pi}}\frac{1}{w^2}.
\end{equation}
If the 
%effective quantum state 
operator $\tilde \rho$ does not show squeezing,  $\langle \Delta x^2 \rangle_{\tilde \rho}\ge 1$, 
the original quantum state $\rho$ can still be squeezed, so that  $\langle \Delta x^2 
\rangle_{\rho} <1$ is possible. The experimental statistical 
uncertainty does not change during the inversion of the moments, therefore one may 
see significant effects after this procedure. For Eq.~\eqref{sq-var} 
this is trivial, but it may also work for more complicated functions 
of moments, such as the Mandel-Q-parameter discussed above. Therefore, it might sometimes be 
useful to calculate normally or standard ordered moments from the 
measured nonclassicality moments.

\subsection{Arbitrary observables}

Let us shortly comment on the expectation value of an arbitrary 
observable. The equations~\eqref{observ} 
and~\eqref{observable} provide the possibility of 
calculating any expectation value in terms of nonclassicality moments. 
Therefore, the knowledge of these moments enables one to obtain 
well-known physical quantities. For instance, the mean energy of the 
harmonic oscillator is given by
\begin{equation}
 \langle  \hat H\rangle = \hbar\omega \left(\langle \hat a^\dagger\hat a\rangle +\frac{1}{2}\right )= 
\hbar\omega\left( M_{\Omega,11} - \sqrt{\frac{2}{\pi}} \frac{1}{w^2} + 
\frac{1}{2}\right),
\end{equation}
where we used Eq.~\eqref{mom11} for the moment $\langle\hat 
a^\dagger\hat a\rangle$.
With the help of Eq.~\eqref{mom22}, a similar, but more lengthy 
expression can be obtained for the variance of the energy 
$\langle(\Delta\hat H)^2\rangle$. By application 
of~Eq.~\eqref{observable}, one can find arbitrary 
expressions for other observables in terms of the accessible nonclassicality moments $M_{\Omega,nm}$.

\section{Summary and Conclusions}\label{conclusions}

We have introduced a method for the verification of nonclassicality of quantum states in terms of the moments of nonclassicality quasiprobabilities. The latter are regularized versions of the Glauber-Sudarshan $P$~function, they display nonclassical effects in terms of negativities of regular functions. Beside the direct visibility of quantum effects as negativities, a strong point is that the nonclassicality quasiprobabilities are available from experimental data by direct sampling. Given the quasiprobabilities, the quantum state of the system is fully characterized and it is straightforward to derive the corresponding moments, called the nonclassicality moments.

It has been demonstrated that all nonclassicality criteria based on normally-ordered moments can also be reformulated in terms of nonclassicality moments. Hence, one can directly formulate the known nonclassicality conditions by replacing therein the normally-ordered moments by the nonclassicality moments. Furthermore, we have derived the relations between the normally-ordered moments and the  the nonclassicality moments. On this basis, one can readily obtain the normally-ordered moments and test the given quantum state based on the standard criteria, without the need to reconstruct normally-ordered moments from the $P$ function. The relations between the moments also enable one to calculate the expectation values of any observable in terms of the nonclassicality moments, and hence directly from the experimentally accessible nonclassicality quasiprobabilities. Finally, we have shown that for infinite filter width the nonclassicality moments converge to the normally-ordered ones. For large values of the width the nonclassicality moments can be regarded as good approximation of the normally-ordered moments. 

The two possibilities for observing nonclassical effects have been illustrated for elementary examples. To analyze sub-Poissonian photon statistics, we 
have studied both the original Mandel-Q-parameter and its counterpart formulated in terms of nonclassicality moments. Both approaches have been applied to a single-photon added thermal state. The possibilities of identifying quadrature squeezing have also be analyzed by using the two kinds of moments. Based on the derived relations between the two types of moments, the nonclassicality quasiprobability  becomes a very general and powerful tool, since it is easily obtained in experiments and one may get very general insight into the physical properties of the quantum state under study.

\section*{Acknowledgement}

The authors gratefully acknowledge valuable comments by Jan Sperling and support by the Deutsche Forschungsgemeinschaft through SFB 652.

%\bibliographystyle{abbrv}
%\bibliography{main}

\end{document}